\documentclass[11pt]{article}

\usepackage[b5paper,hmargin=20mm,top=22mm,bottom=25mm]{geometry}
\usepackage{amsmath,amssymb,amsfonts}
\usepackage{mathtools}
\usepackage{bm}
\usepackage{upgreek}
\usepackage{graphicx}
\usepackage[font=small,labelfont=bf,labelsep=period,figurename=Fig.]{caption}
\usepackage{placeins}
\usepackage{xcolor}
\usepackage{authblk}

\usepackage[numbers,sort&compress]{natbib}
\usepackage[hidelinks]{hyperref}

\newcommand{\abs}[1]{\left\lvert#1\right\rvert}

\newcommand{\Da}{\mathrm{Da}_s}
\newcommand{\Davi}{\mathrm{Da}_v}

\title{Energetic costs of reaching the physical limits of sensing}

\author[1]{Albert Alonso}
\author[2]{Julius B. Kirkegaard}
\author[3]{Robert G. Endres}

\affil[1]{Department of Computer Science, Aalborg University, Copenhagen, Denmark}
\affil[2]{Department of Computer Science \& Niels Bohr Institute, University of Copenhagen, Copenhagen, Denmark}
\affil[3]{Department of Life Sciences, and Centre for Integrative Systems Biology and Bioinformatics, Imperial College London, London, United Kingdom}

\date{}

\begin{document}

\maketitle

\begin{abstract}
Cells sense chemical gradients by absorbing ligand molecules, a nonequilibrium process that necessarily produces entropy.
But how many molecules must a cell absorb, and therefore energy dissipate, to approach the physical limits of sensing?
We derive closed-form expressions relating gradient sensing uncertainty to the number of absorbed molecules, which we link to entropy production.
Unlike concentration sensing, which depends only on the total count, gradient sensing must resolve the angular asymmetry of the absorbed flux, adding a bounded dipole-contrast factor set by how strongly the cell absorbs.
In the strong-signal regime, cells approach peak chemotactic alignment while absorbing a small fraction of the diffusion-limited uptake, with the required number of uptake events set by the relative concentration change across the cell.
Re-analysis of \textit{Dictyostelium} chemotaxis places the cell at $0.25\%$ of that limit with a minor metabolic load, suggesting that sensing is limited by counting noise rather than by energetic cost.
\end{abstract}

Cells navigate chemical landscapes by sensing concentration gradients through surface receptors, a process central to immune response, embryonic development, and microbial foraging~\cite{Swaney2010}.
Accurate sensing requires sampling enough ligand molecules to overcome the intrinsic stochasticity of diffusive encounters~\cite{Aquino2016}.
This in turn demands metabolic resources, as receptors must be synthesized, ligands internalized or degraded, and intracellular signaling cascades maintained out of equilibrium~\cite{Lan2012}.

\citet{Berg1977} established the fundamental precision of concentration sensing, showing that the relative uncertainty of a cell of radius $a$ estimating a background concentration $c$ over time $T$ scales as $1/\sqrt{DacT}$, with $D$ being the ligand diffusion constant.
This bound reflects the counting noise of diffusive molecular encounters.
\citet{Endres2008} formalized the perfect absorber, a sphere that irreversibly internalizes every arriving ligand, suppressing the re-encounters that correlate successive observations at a reversibly binding surface so that every count is an independent sample. It is the canonical realization of this limit, and they extended it to spatial gradient sensing, where the same diffusive pre-factor governs the estimation of $\nabla c$.
Further work addressed temporal sensing~\cite{Mora2010}, sensor geometry and domain size~\cite{McCusker2025}, and the role of receptor
correlations~\cite{Kaizu2014a, Alonso2025b}.

A separate line of work has addressed the thermodynamic cost of precise sensing.
Biochemical information processing carries energetic costs,
with cellular computation constrained by tradeoffs between accuracy,
speed, and dissipation \cite{Mehta2012, Lan2012}.
More broadly, the thermodynamic uncertainty relation (TUR) \cite{Barato2015} provides a general bound linking the precision of any nonequilibrium current to the entropy production \cite{Horowitz2020}.
\citet{Lang2014} showed that nonequilibrium driving of receptor kinetics can push sensing accuracy from the Berg--Purcell value toward the maximum-likelihood limit, and \citet{Harvey2023} applied the TUR framework to quantify energy--accuracy tradeoffs at the single-receptor level.
However, these analyses operate at the scale of individual molecular reactions, leaving open how thermodynamic costs translate to whole-cell gradient sensing and ultimately to behavioral phenomena such as cellular chemotaxis~\cite{Alonso2024}.

\begin{figure}[htbp]
\centering
\includegraphics[width=.7\linewidth]{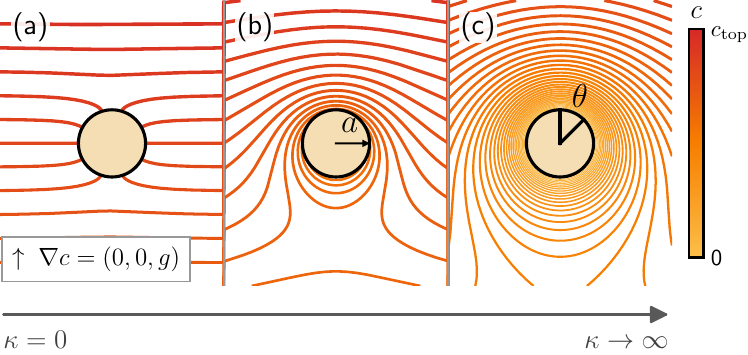}
\caption{\textbf{Gradient sensing by a partially absorbing sphere of radius~$a$ in a linear concentration gradient.} The surface reactivity $\kappa$ ranges from (a)~a perfect reflector ($\kappa=0$), to (b)~a partial absorber ($0<\kappa<\infty$), to (c)~a perfect absorber ($\kappa\to\infty$).
Curves are iso-concentration lines, colored and weighted by $c$, and $\theta$ is measured from $\nabla c$.}
\label{fig:fig1}
\end{figure}

Here, we bridge this gap by studying a partially absorbing cell that interpolates continuously between a perfect reflector, which is in equilibrium with its surroundings and gains no information, and the perfect absorber~(Fig.~\ref{fig:fig1}).
By deriving sensing accuracy and entropy production analytically as functions of the absorption strength, we quantify the tradeoff between precision and energetic expenditure, and connect it to chemotactic performance.
Our results clarify how close biological sensing operates to fundamental thermodynamic bounds and reveal the regime in which energetic investment yields diminishing returns.

\paragraph{Model ---}
We model a cell of radius $a$ as a partially absorbing sphere in a ligand field by imposing a mixed (Robin) boundary condition on the steady-state diffusion problem (Fig.~\ref{fig:fig1}),
\begin{equation}
D\,\partial_r c\big|_{r=a}=\kappa\,c(a,\theta),
\label{eq:robin}
\end{equation}
with diffusion constant $D$ and surface reactivity $\kappa$ (inward flux convention).
The dimensionless control parameter is the surface Damk\"ohler number
\begin{equation}\label{eq:damkohler}
\mathrm{Da}_s\equiv \frac{\kappa a}{D},
\end{equation}
which interpolates between a perfect reflector ($\mathrm{Da}_s\to 0$) and a perfect absorber ($\mathrm{Da}_s\to\infty$).
While the spherical geometry is convenient, the relations below carry over to a cell adhered to a planar substrate, and are only moderately altered by departures from sphericity (see Appendix~\ref{app:geometry}).

\paragraph{Monopole--dipole decomposition and gradient-sensing limit ---}
To quantify how partial uptake reshapes gradient sensing, we decompose the surface flux into its angular modes.
In a shallow gradient along the $z$ axis, i.e.\ ${ag/c_0\ll1}$ for all conditions analyzed below (see Appendix~\ref{app:gradient}), the far-field concentration is ${c(r,\theta)\to c_0+g\,r\cos\theta}$ and the surface flux takes the form:
\begin{equation}
j(\theta)=j_0+j_1\cos\theta,
\label{eq:flux_decomp}
\end{equation}
where $j_0$ is the isotropic (monopole) component and $j_1$ the dipolar front--back asymmetry.
Solving Laplace's equation with Eq.~\eqref{eq:robin} yields
\begin{equation}
j_0=\frac{D\kappa c_0}{D+\kappa a},
\qquad
j_1=\frac{3D\kappa a\, g}{2D+\kappa a},
\label{eq:j0j1}
\end{equation}
and therefore the key ratio 
\begin{equation}
\frac{j_1}{j_0}=\frac{3ag}{c_0}\cdot\frac{1+\mathrm{Da}_s}{2+\mathrm{Da}_s}.
\label{eq:ratio}
\end{equation}

Equation~\eqref{eq:ratio} shows that partial absorption changes the
angular contrast of the absorbed flux only through the bounded factor
$(1+\mathrm{Da}_s)/(2+\mathrm{Da}_s)$, which increases from $1/2$ to
unity between the reflecting and absorbing limits.
However, this ratio alone does not determine the available sensing
information, because both $j_0$ and $j_1$ vanish as
$\mathrm{Da}_s\to0$.
Thus, while a weakly absorbing sphere may have a well-defined formal
monopole--dipole ratio, it absorbs too few molecules to estimate the
gradient.

The total number of absorbed ligands in time $T$ follows from the monopole flux,
\begin{equation}
\langle N\rangle = 4\pi a^2 j_0\,T
= 4\pi D a c_0 T\,\frac{\mathrm{Da}_s}{1+\mathrm{Da}_s},
\label{eq:N}
\end{equation}
which returns the~\citet{Berg1977} perfect-absorber limit as $\mathrm{Da}_s\to\infty$.

\paragraph{Gradient sensing uncertainty ---}
To estimate the gradient, the cell must distinguish how many ligands arrive at the front versus the back of its surface.
The natural statistic is the flux-weighted dipole moment
${\langle\cos\theta\rangle_j=j_1/(3j_0)}$,
proportional to the gradient $g$ by construction.
Under Poisson absorption statistics, the squared relative uncertainty of the resulting estimator factorizes as (see Appendix~\ref{app:gradient})
\begin{equation}
\left(\frac{\delta g}{g}\right)^2
=
\frac{3}{\langle N\rangle}
\left(\frac{j_0}{j_1}\right)^2,
\label{eq:grad_uncertainty}
\end{equation}
where the first factor is the counting noise, shared with concentration sensing and reduced by absorbing more ligand, and the second reflects the difficulty of extracting a weak dipole from a dominant monopole background.
Expressing the gradient error in units of $c_0/a$ removes the explicit signal prefactor in Eq.~\eqref{eq:ratio}.
Substituting Eqs.~\eqref{eq:j0j1}--\eqref{eq:N} then yields the closed-form physical limit
\begin{equation}
\left(\frac{\delta g}{c_0/a}\right)^2
=
\frac{1}{12\pi D a c_0 T}\,
\frac{(2+\mathrm{Da}_s)^2}{\mathrm{Da}_s(1+\mathrm{Da}_s)},
\label{eq:grad_uncertainty_scaled}
\end{equation}
which recovers the perfect-absorber result of~\citet{Endres2008} as $\mathrm{Da}_s\to\infty$, and diverges as $\mathrm{Da}_s\to0$ as uptake vanishes.

Equation~\eqref{eq:grad_uncertainty_scaled} can also be written in terms of counted molecules alone.
Introducing the uptake fraction
\begin{equation}
\Phi\equiv\frac{\langle N\rangle}{\langle N\rangle_{\max}}=\frac{\mathrm{Da}_s}{1+\mathrm{Da}_s},
\qquad
\langle N\rangle_{\max}=4\pi Dac_0T,
\label{eq:n-def}
\end{equation}
Eq.~\eqref{eq:grad_uncertainty_scaled} becomes exactly
\begin{equation}
\left(\frac{\delta g}{c_0/a}\right)^2
=\frac{\left(2-\Phi\right)^2}{3\,\langle N\rangle}.
\label{eq:counting}
\end{equation}
Beyond the count itself, the only dependence left is the bounded dipole-contrast factor ${(2-\Phi)^2}$, which falls from $4$ at the reflector to $1$ at the perfect absorber.
It stays within a few percent of $4$ at every uptake we infer below, so the penalty is effectively constant away from the diffusion limit.
Notably, concentration sensing carries no such factor, and its uncertainty is the counting noise alone.

\paragraph{Energetic cost  ---}
Each absorbed ligand is internalized or degraded, and such a nonequilibrium process must produce entropy.
Over a measurement of duration $T$, the entropy produced is
\begin{equation}
\Sigma=\langle N\rangle\left[\Delta s_0-\ln(1-\Phi)\right],
\label{eq:Sigma}
\end{equation}
with $\Delta s_0$ the entropy produced per uptake event by the internal signaling machinery, and the second term the contribution of transporting the ligand from the bulk to the cell surface (see Appendix~\ref{app:entropy-derivation}).

The uptake cost $\Delta s_0$ represents every driven, nearly irreversible step by which a capture is registered and amplified, from the receptor cycle itself to the downstream reactions that hold the signaling network out of equilibrium.
Its magnitude is therefore architecture-dependent.
G-protein-coupled amplification~\cite{Janetopoulos2001} places it at $10^2$--$10^3\,k_B$, each nucleotide hydrolysis costing $\sim\!20\,k_B$~\cite{Milo2015}, whereas optimal resource allocation puts the cost of an independent measurement near $1\,k_B$~\cite{Govern2014a, Govern2014b, TenWolde2016}, an accounting per receptor-occupancy sample rather than per absorbed molecule.
We therefore leave $\Delta s_0$ as a free parameter.

\begin{figure}[htbp]
    \centering
    \includegraphics[width=.75\linewidth]{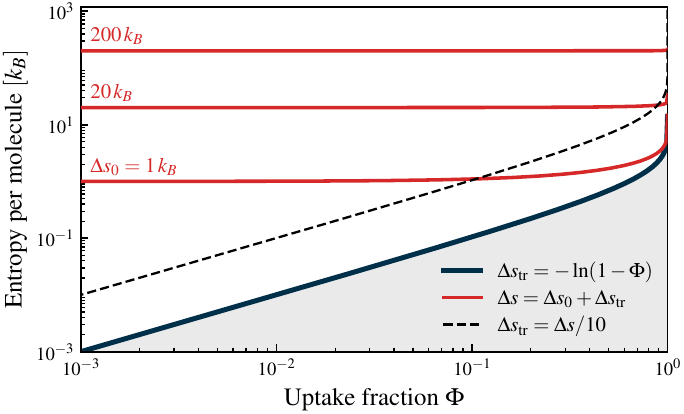}
    \caption{
    \textbf{Entropy per absorbed molecule.}
    The transport term $\Delta s_{\mathrm{tr}}=-\ln(1-\Phi)$, Eq.~\eqref{eq:transport-integral},
    is unavoidable, so the shaded region below it is inaccessible.
    Each red curve is the total ${\Delta s=\Delta s_0+\Delta s_{\mathrm{tr}}}$ for a signaling
    cost $\Delta s_0=1$, $20$ or $200\,k_B$, and stays at its $\Delta s_0$ until close to the
    diffusion limit.
    The dashed locus, ${\Phi=1-e^{-\Delta s_0/9}}$, marks where transport reaches a tenth of
    the total.
    }
    \label{fig:fig3}
\end{figure}

The transport term follows from the concentration profile alone and represents the entropy produced in the medium as a ligand diffuses from the bulk into the region the cell has depleted.
In the steady state the local dissipation of ideal diffusion is ${\sigma=D\abs{\nabla c}^2/c}$.
Integrating it over the exterior with $u=r/a$ and dividing by the uptake current $4\pi Dac_0\Phi$ gives the entropy produced per absorbed molecule,
\begin{equation}
\Delta s_{\mathrm{tr}}=\Phi\!\int_1^\infty\!\frac{du}{u(u-\Phi)}=-\ln(1-\Phi)=\ln\frac{c_0}{c(a)},
\label{eq:transport-integral}
\end{equation}
the chemical-potential drop each ligand falls through on its way to the surface, effectively a binding free energy with the surface concentration in place of a dissociation constant (see Appendix~\ref{app:transport} for a detailed derivation).
Notably, it vanishes at the reflector, where the cell is in equilibrium with its surroundings, and diverges logarithmically at the perfect absorber, where $c(a)\to0$ (Fig.~\ref{fig:fig3}).
Such transport entropy is independent of the cell's shape: for any body held at a uniform surface concentration the same co-area argument returns Eq.~\eqref{eq:transport-integral}.

No finite entropy per molecule therefore attains the diffusion limit.
\citet{Lang2014} reached the same conclusion for maximum-likelihood estimation, where the divergence arises in the driven receptor kinetics rather than in transport.
It is logarithmic, however, and therefore slow.
With $\Delta s_0\simeq20\,k_B$ for a single GTP hydrolysis~\cite{Milo2015}, transport supplies less than $10\%$ of the total for all $\Phi<0.89$, and the two terms reach parity only within $2\times10^{-9}$ of the diffusion limit (Fig.~\ref{fig:fig3}).
Larger $\Delta s_0$ pushes that crossover closer still.
Away from the perfect absorber the bracket in Eq.~\eqref{eq:Sigma} is therefore $\Delta s_0$ to within a few percent, so the dissipation is proportional to the number absorbed.

Eliminating $\langle N\rangle$ between Eqs.~\eqref{eq:counting} and~\eqref{eq:Sigma} makes the counting relation an exact energy--accuracy relation, which discarding the transport term, ${-\ln(1-\Phi)\geq0}$, turns into a bound (see Appendix~\ref{app:entropy-derivation}),
\begin{align}
\left(\frac{\delta g}{c_0/a}\right)^2
&=\frac{(2-\Phi)^2\left[\Delta s_0-\ln(1-\Phi)\right]}{3\,\Sigma}\nonumber\\[2pt]
&\geq\;\frac{(2-\Phi)^2\Delta s_0}{3\,\Sigma}.
\label{eq:tradeoff}
\end{align}
The bound retains $\Delta s_0$ alone, the biochemical cost of registering a capture, and discards the transport term, which is set by the cell's uptake but dissipated in the surrounding solution.
It holds at every uptake fraction, and is tight to within ${\Delta s_{\mathrm{tr}}/\Delta s_0}$, so it is tight wherever transport is cheap against the signaling cost (Fig.~\ref{fig:fig3}).
Since $\Delta s_0$ omits receptor synthesis and the upkeep of the network, the bound is also a conservative statement about metabolic cost.

Equation~\eqref{eq:tradeoff} has the form of a thermodynamic uncertainty relation~\cite{Barato2015, Horowitz2020}, the squared relative error falling as the inverse of the dissipation, but the prefactor is fixed by the signaling architecture through $\Delta s_0$ rather than by a universal constant.

\paragraph{Macroscopic effects of energetic cost ---}

To relate sensing accuracy to behavioral output, we study how the absorption–dissipation tradeoff propagates to the chemotactic index (CI), defined as the mean alignment between cell velocity and gradient direction.
Following~\cite{Endres2008, Alonso2025a}, CI takes the closed form (see Appendix~\ref{app:ci})
\begin{equation}
\mathrm{CI}(z)=\sqrt{\tfrac{\pi z}{2}}\,e^{-z}\,\big[I_0(z)+I_1(z)\big],
\label{eq:CI}
\end{equation}
with $I_0$ and $I_1$ the modified Bessel functions and ${z=(g/\delta g)^2/4}$.
By Eq.~\eqref{eq:counting},
\begin{equation}
    z = A\,\frac{\Phi}{\left(2-\Phi\right)^2}, \qquad A \equiv 3\pi D a^3 T\,\frac{g^2}{c_0},
    \label{eq:z-sigma}
\end{equation}
so that the entire behavioral response depends on the environment and sensing time only through the single dimensionless group $A$ (see Appendix~\ref{app:units} for the dimensional analysis) and on the uptake fraction $\Phi$.
As $g^2/c_0$ plays the role of a signal-to-noise ratio (SNR), this reproduces the scaling collapse found in~\cite{Endres2008, Alonso2025a}.

\begin{figure}[htbp]
    \centering
    \includegraphics[width=.75\linewidth]{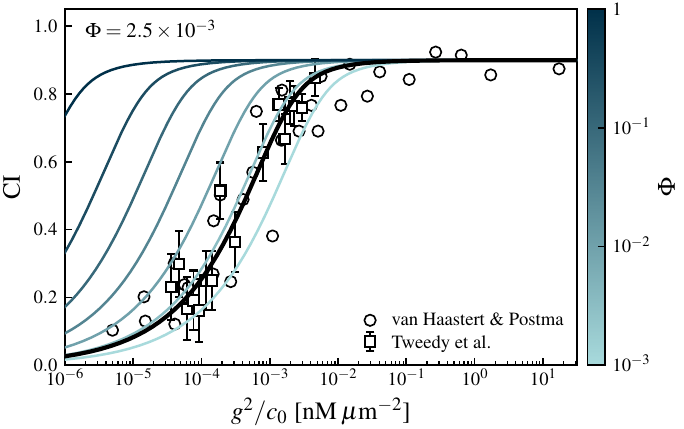}
    \caption{\textbf{Chemotactic index of \textit{D.~discoideum} against the fitted model.}
    CI against $g^2/c_0$ (nM\,$\upmu$m$^{-2}$, converted with
    $1$\,nM\,$=0.6022\;\upmu$m$^{-3}$) for a range of uptake fractions $\Phi$ (color), at
    $Da^3T=1.5\times10^{5}\;\upmu$m$^5$. Symbols are the data
    of~\cite{VanHaastert2007, Tweedy2013}, the solid line the best fit,
    $\Phi=2.5\times10^{-3}$, with the empirical prefactor $0.9$ of~\cite{Endres2008}.}
    \label{fig:chemo}
\end{figure}

Figure~\ref{fig:chemo} compares Eq.~\eqref{eq:CI} with chemotactic data from \textit{Dictyostelium discoideum}~\cite{VanHaastert2007, Tweedy2013}, with the empirical prefactor $0.9$ of~\cite{Endres2008} accounting for the fact that not all cells perform chemotaxis.
Based on the $2$--$8$~s time-averaging window that~\citet{VanHaastert2007} infer from receptor noise, we set $T=4$~s, alongside $D=300\;\upmu$m$^2$/s and $a=5\;\upmu$m~\cite{Endres2008}, making the uptake fraction the only free parameter, whose fit yields
\begin{equation}
\Phi_{\mathrm{Dicty}}=(2.5\pm0.3)\times10^{-3},
\label{eq:n-dicty}
\end{equation}
corresponding to a surface reactivity $\kappa\simeq0.15\;\upmu$m/s.
\textit{Dictyostelium} thus absorbs about $0.25\%$ of the diffusion-limited uptake.
This benefit of absorption extends to reversible ligand--receptor binding with indiscriminate degradation of ligand at the surface, as utilized by \textit{Dictyostelium} through its receptor cAR1 and the extracellular phosphodiesterase PdsA~\cite{VanHaastert1984, Endres2008}.
Across their threshold conditions this is tens to hundreds of ligands, $\langle N\rangle\simeq810$ at the most concentrated.
This count is insensitive to the assumed $T$.
Across the $2$--$8$~s window the fitted $\Phi$ ranges from $5.0\times10^{-3}$ to $1.3\times10^{-3}$, but since $\langle N\rangle_{\max}\propto T$, $\langle N\rangle$ stays between $809$ and $813$ ligands.
At this uptake the transport term of Eq.~\eqref{eq:Sigma} is only $2.5\times10^{-3}\,k_B$ per molecule, so ${\Sigma\simeq\langle N\rangle\Delta s_0}$ and the dissipation is proportional to the count.

\begin{figure}[htbp]
    \centering
    \includegraphics[width=.75\linewidth]{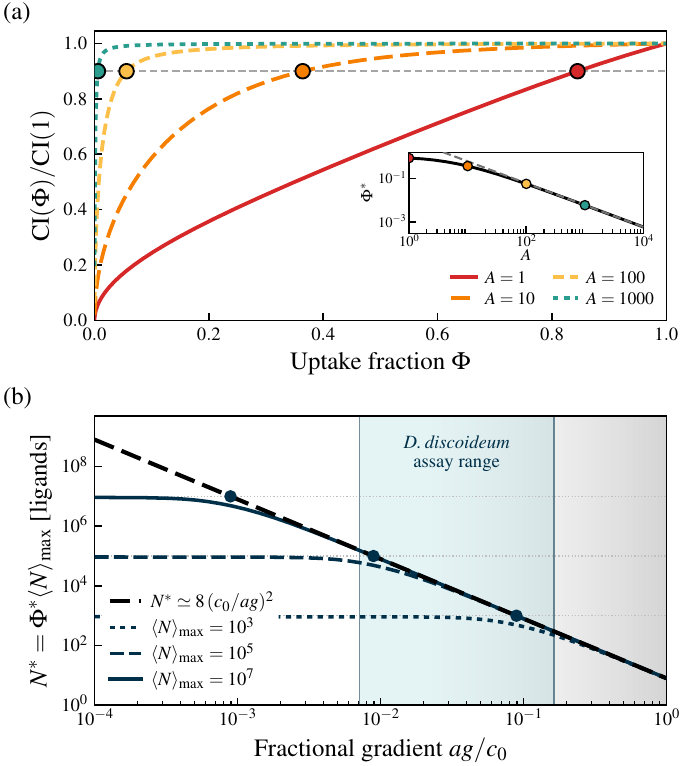}
    \caption{\textbf{Uptake required for near-peak chemotaxis.}
    (a)~Normalized response $\mathrm{CI}(\Phi)/\mathrm{CI}(1)$ for four values of
    $A=3\pi Da^3T\,g^2/c_0$, with $\alpha=0.9$ dashed and the corresponding $\Phi^\ast$
    circled. Inset, $\Phi^\ast(A)$ and its asymptote $4\,\mathrm{CI}^{-1}(\alpha)/A$.
    (b)~Ligand count for $\alpha=0.9$ against $ag/c_0$ at
    $\langle N\rangle_{\max}=10^3$, $10^5$ and $10^7$, circles at their ceilings, the
    asymptote $N^\ast\simeq8\,(c_0/ag)^2$ of Eq.~\eqref{eq:sigmastar-asymp} dashed. The band is the
    \textit{D.~discoideum} threshold range of~\citet{VanHaastert2007}, $0.7\%$ to $16\%$,
    grey the growth of shallow-gradient corrections.}
    \label{fig:cost}
\end{figure}

We now ask how much uptake is required to approach peak CI.
Figure~\ref{fig:cost}(a) shows ${\mathrm{CI}(z)/\mathrm{CI}(z_{\max})}$ against $\Phi$, with ${z_{\max}=z(\Phi=1)=A}$ its perfect-absorber value.
For ${A\gtrsim10^2}$ the response saturates well before ${\Phi=1}$.
Note that a large $A$ can be achieved by a cell in a shallow gradient by integrating for longer $T$.
Defining $\Phi^*$ as the smallest uptake fraction, and hence the least dissipative, that yields ${\mathrm{CI}=\alpha\,\mathrm{CI}(z_{\max})}$, inversion of Eq.~\eqref{eq:z-sigma} gives the closed form
\begin{equation}
    \Phi^\ast \;=\; \frac{1 + 4\zeta - \sqrt{1 + 8\zeta}}{2\zeta},
    \label{eq:ustar}
\end{equation}
with ${\zeta \equiv \frac{1}{A}\mathrm{CI}^{-1}\!\big(\alpha\,\mathrm{CI}(A)\big)}$.
In the strong-signal regime (${A \gg 1}$), ${\zeta \to 0}$ and Eq.~\eqref{eq:ustar} reduces to ${\Phi^\ast \approx 4\,\mathrm{CI}^{-1}(\alpha)/A}$, which in absolute terms becomes
\begin{equation}
N^* \;=\; \Phi^*\langle N\rangle_{\max} \;\approx\; \frac{16}{3}\,\mathrm{CI}^{-1}(\alpha)\left(\frac{c_0}{ag}\right)^2 .
\label{eq:sigmastar-asymp}
\end{equation}
Interestingly, the diffusion constant and the integration time cancel out, so the cost-optimal number of absorbed ligands is set by the fractional gradient $ag/c_0$ alone.
Since the dependence is inverse-square, halving the concentration change across the cell quadruples the uptake that near-peak chemotaxis demands.
For ${\alpha=0.9}$, the \textit{Dictyostelium} assays span from a few hundred ligands in their steepest gradients to about $10^5$ in their shallowest (Fig.~\ref{fig:cost}(b)).
Note that these are counts for near-peak alignment, not for detection, which is the condition the assays report.
The uptake fractions they call for run from a few percent to about a third of the diffusion limit, where transport still contributes under $0.4\,k_B$ per molecule, so ${\Sigma\simeq\langle N\rangle\Delta s_0}$ throughout.

How realistic are these uptakes in the context of a cell's overall metabolic budget? A chemotactic \textit{Dictyostelium discoideum} cell operates at $\sim 10^8\text{--}10^9$ ATP/s, with only a small fraction devoted to receptor-level sensing ($\sim 1\text{--}5\%$) \cite{Milo2015, Zanotelli2018}. Even at the GPCR-amplified end of the range, ${\Delta s_0\sim10^2\text{--}10^3\,k_B}$ per uptake event, the $\sim\!10^3$ ligands \textit{Dictyostelium} absorbs cost $\sim\!5\times10^{4}$ ATP over the measurement, at worst about $1\%$ of that allowance. What limits sensing at this uptake is therefore the number of molecules counted, not the energy spent.
Since the cost is energetically negligible, what bounds uptake is likely to be its other consequences, such as depleting a signal that neighboring cells also rely on.
Indeed, in \textit{Dictyostelium} both receptor cAR1 and phosphodiesterase PdsA are developmentally upregulated during starvation and aggregation~\cite{Louis1993, Masaki2013}.

In this Letter, we derived a closed-form relation between the number of absorbed molecules and the accuracy of whole-cell gradient sensing, and hence between accuracy and entropy production.
Unlike the classical Berg--Purcell picture, where accuracy improves through increased sampling alone, gradient sensing also carries a dipole-contrast penalty from resolving angular asymmetry in the absorbed flux.
The entropy produced per absorbed molecule is bounded below by the chemical-potential drop the ligand falls through, $\ln[c_0/c(a)]$, which diverges at the diffusion limit, so that limit cannot be reached at finite dissipation.
Propagating the relation to behavior yields a closed-form threshold for near-peak chemotaxis, showing that in the strong-signal regime the required number of uptake events is independent of the diffusion constant and integration time.

The \textit{Dictyostelium} data place the cell far below the perfect-absorber limit, at ${\Phi\simeq2.5\times10^{-3}}$ of the diffusion-limited uptake, with tens to hundreds of absorbed ligands across the chemotaxis thresholds assayed.
More generally, we show that the perfect-absorber limit is not only unreachable but often also unnecessary, as a small fraction of the diffusion-limited uptake suffices for near-peak chemotaxis.
This diminishing return in absorption-based concentration sensing is already hinted at by \citet{Berg1977} when hypothesizing about a finite number of absorbing patches, and here we confirm it and show how it relates to chemotaxis and gradient sensing more generally.
The theory therefore predicts that cells in stronger fractional gradients should operate farther below the perfect-absorber limit, whereas those in weaker ones may require uptake closer to the diffusion-limited regime, a third of it already at the shallowest gradient assayed here.
Since signal also scales with the integration time, and that of a freely swimming bacterium is bounded by rotational diffusion and shorter than a crawling eukaryote's, bacteria should need a larger fraction still of the diffusion-limited uptake (see Appendix~\ref{app:ci}).
Similarly and not explored here is the effect of not operating in the absorber limit in a collective environment, where strong uptake, while suppressing rebinding and improving single-cell sensing, depletes the shared diffusive environment affecting its neighbors~\cite{Kirkegaard2026}.

\section*{Acknowledgments}
This work was supported by the Novo Nordisk Foundation under Grant Agreement NNF20OC0062047 for J.B.K. and the AI-4-EB Consortium for Bioengineered Cells and Systems (BBSRC award BB/W013770/1) for support to R.G.E.

\FloatBarrier
\bibliography{references}

\begin{thebibliography}{29}
\providecommand{\natexlab}[1]{#1}
\providecommand{\url}[1]{\texttt{#1}}
\expandafter\ifx\csname urlstyle\endcsname\relax
  \providecommand{\doi}[1]{doi: #1}\else
  \providecommand{\doi}{doi: \begingroup \urlstyle{rm}\Url}\fi

\bibitem[Swaney et~al.(2010)Swaney, Huang, and Devreotes]{Swaney2010}
K.F. Swaney, C.-H. Huang, and P.N. Devreotes.
\newblock Eukaryotic chemotaxis: A network of signaling pathways controls
  motility, directional sensing, and polarity.
\newblock \emph{Annual Review of Biophysics}, 39:\penalty0 265--289, 2010.
\newblock \doi{10.1146/annurev-biophys-070909-104804}.

\bibitem[Aquino et~al.(2016)Aquino, Wingreen, and Endres]{Aquino2016}
G.~Aquino, N.S. Wingreen, and R.G. Endres.
\newblock Know the single-receptor sensing limit? think again.
\newblock \emph{Journal of Statistical Physics}, 162\penalty0 (5):\penalty0
  1353--1364, 2016.
\newblock \doi{10.1007/s10955-015-1412-9}.

\bibitem[Lan et~al.(2012)Lan, Sartori, Neumann, Sourjik, and Tu]{Lan2012}
G.~Lan, P.~Sartori, S.~Neumann, V.~Sourjik, and Y.~Tu.
\newblock The energy--speed--accuracy trade-off in sensory adaptation.
\newblock \emph{Nature Physics}, 8:\penalty0 422--428, 2012.
\newblock \doi{10.1038/nphys2276}.

\bibitem[Berg and Purcell(1977)]{Berg1977}
Howard~C. Berg and Edward~M. Purcell.
\newblock Physics of chemoreception.
\newblock \emph{Biophysical Journal}, 20\penalty0 (2):\penalty0 193--219, 1977.
\newblock \doi{10.1016/S0006-3495(77)85544-6}.

\bibitem[Endres and Wingreen(2008)]{Endres2008}
Robert~G. Endres and Ned~S. Wingreen.
\newblock Accuracy of direct gradient sensing by single cells.
\newblock \emph{Proceedings of the National Academy of Sciences USA},
  105\penalty0 (41):\penalty0 15749--15754, 2008.
\newblock \doi{10.1073/pnas.0804686105}.

\bibitem[Mora and Wingreen(2010)]{Mora2010}
Thierry Mora and Ned~S. Wingreen.
\newblock Limits of sensing temporal concentration changes by single cells.
\newblock \emph{Physical Review Letters}, 104\penalty0 (24):\penalty0 248101,
  2010.
\newblock \doi{10.1103/PhysRevLett.104.248101}.

\bibitem[McCusker and Lubensky(2025)]{McCusker2025}
D.~R. McCusker and D.~K. Lubensky.
\newblock Physical limits on chemical sensing in bounded domains.
\newblock \emph{Physical Review E}, 111\penalty0 (3):\penalty0 034404, 2025.
\newblock \doi{10.1103/PhysRevE.111.034404}.

\bibitem[Kaizu et~al.(2014)Kaizu, de~Ronde, Paijmans, Takahashi, Tostevin, and
  ten Wolde]{Kaizu2014a}
Kazunari Kaizu, Wouter de~Ronde, Joris Paijmans, Kazuhiro Takahashi, Filipe
  Tostevin, and Pieter~Rein ten Wolde.
\newblock The {B}erg-{P}urcell limit revisited.
\newblock \emph{Biophysical Journal}, 106\penalty0 (4):\penalty0 976--985,
  2014.
\newblock \doi{10.1016/j.bpj.2013.12.030}.

\bibitem[Alonso et~al.(2025{\natexlab{a}})Alonso, Endres, and
  Kirkegaard]{Alonso2025b}
Albert Alonso, Robert~G. Endres, and Julius~B. Kirkegaard.
\newblock Local clustering and global spreading of receptors for optimal
  spatial gradient sensing.
\newblock \emph{Physical Review Letters}, 134\penalty0 (15):\penalty0 158401,
  2025{\natexlab{a}}.
\newblock \doi{10.1103/PhysRevLett.134.158401}.

\bibitem[Mehta and Schwab(2012)]{Mehta2012}
Pankaj Mehta and David~J Schwab.
\newblock Energetic costs of cellular computation.
\newblock \emph{Proceedings of the National Academy of Sciences}, 109\penalty0
  (44):\penalty0 17978--17982, 2012.

\bibitem[Barato and Seifert(2015)]{Barato2015}
Andre~C. Barato and Udo Seifert.
\newblock Thermodynamic uncertainty relation for biomolecular processes.
\newblock \emph{Physical Review Letters}, 114\penalty0 (15):\penalty0 158101,
  2015.
\newblock \doi{10.1103/PhysRevLett.114.158101}.

\bibitem[Horowitz and Gingrich(2020)]{Horowitz2020}
Jordan~M. Horowitz and Todd~R. Gingrich.
\newblock Thermodynamic uncertainty relations constrain non-equilibrium
  fluctuations.
\newblock \emph{Nature Physics}, 16:\penalty0 15--20, 2020.
\newblock \doi{10.1038/s41567-019-0702-6}.

\bibitem[Lang et~al.(2014)Lang, Fisher, Mora, and Mehta]{Lang2014}
Alex~H. Lang, Charles~K. Fisher, Thierry Mora, and Pankaj Mehta.
\newblock Thermodynamics of statistical inference by cells.
\newblock \emph{Physical Review Letters}, 113\penalty0 (14):\penalty0 148103,
  2014.
\newblock \doi{10.1103/PhysRevLett.113.148103}.

\bibitem[Harvey et~al.(2023)Harvey, Lahiri, and Ganguli]{Harvey2023}
Sam~E. Harvey, Sayak Lahiri, and Surya Ganguli.
\newblock Universal energy-accuracy tradeoffs in nonequilibrium cellular
  sensing.
\newblock \emph{Physical Review E}, 108\penalty0 (1):\penalty0 014403, 2023.
\newblock \doi{10.1103/PhysRevE.108.014403}.

\bibitem[Alonso and Kirkegaard(2024)]{Alonso2024}
Albert Alonso and Julius~B. Kirkegaard.
\newblock Learning optimal integration of spatial and temporal information in
  noisy chemotaxis.
\newblock \emph{PNAS Nexus}, 3\penalty0 (7):\penalty0 235, 2024.

\bibitem[Janetopoulos et~al.(2001)Janetopoulos, Jin, and
  Devreotes]{Janetopoulos2001}
Christopher Janetopoulos, Tian Jin, and Peter Devreotes.
\newblock Receptor-mediated activation of heterotrimeric {G}-proteins in living
  cells.
\newblock \emph{Science}, 291\penalty0 (5512):\penalty0 2408--2411, 2001.
\newblock \doi{10.1126/science.1055835}.

\bibitem[Milo and Phillips(2015)]{Milo2015}
Ron Milo and Rob Phillips.
\newblock \emph{Cell Biology by the Numbers}.
\newblock Garland Science, 2015.
\newblock ISBN 9780815345374.

\bibitem[Govern and ten Wolde(2014{\natexlab{a}})]{Govern2014a}
Christopher~C Govern and Pieter~Rein ten Wolde.
\newblock Energy dissipation and noise correlations in biochemical sensing.
\newblock \emph{Physical Review Letters}, 113\penalty0 (25):\penalty0 258102,
  2014{\natexlab{a}}.

\bibitem[Govern and ten Wolde(2014{\natexlab{b}})]{Govern2014b}
Christopher~C Govern and Pieter~Rein ten Wolde.
\newblock Optimal resource allocation in cellular sensing systems.
\newblock \emph{Proceedings of the National Academy of Sciences USA},
  111\penalty0 (49):\penalty0 17486--17491, 2014{\natexlab{b}}.

\bibitem[ten Wolde et~al.(2016)ten Wolde, Becker, Ouldridge, and
  Mugler]{TenWolde2016}
Pieter~Rein ten Wolde, Nils~B Becker, Thomas~E Ouldridge, and Andrew Mugler.
\newblock Fundamental limits to cellular sensing.
\newblock \emph{Journal of Statistical Physics}, 162\penalty0 (5):\penalty0
  1395--1424, 2016.

\bibitem[Alonso et~al.(2025{\natexlab{b}})Alonso, Kirkegaard, and
  Endres]{Alonso2025a}
A.~Alonso, J.~B. Kirkegaard, and R.~G. Endres.
\newblock Persistent pseudopod splitting is an effective chemotaxis strategy in
  shallow gradients.
\newblock \emph{Proceedings of the National Academy of Sciences USA},
  122\penalty0 (19):\penalty0 e2502368122, 2025{\natexlab{b}}.
\newblock \doi{10.1073/pnas.2502368122}.

\bibitem[van Haastert and Postma(2007)]{VanHaastert2007}
Peter J.~M. van Haastert and Marten Postma.
\newblock Biased random walk by stochastic fluctuations of
  chemoattractant-receptor interactions at the lower limit of detection.
\newblock \emph{Biophysical Journal}, 93\penalty0 (5):\penalty0 1787--1796,
  2007.
\newblock ISSN 0006-3495.
\newblock \doi{10.1529/biophysj.107.104356}.

\bibitem[Tweedy et~al.(2013)Tweedy, Meier, Stephan, Heinrich, and
  Endres]{Tweedy2013}
Luke Tweedy, Börn Meier, Jürgen Stephan, Doris Heinrich, and Robert~G.
  Endres.
\newblock Distinct cell shapes determine accurate chemotaxis.
\newblock \emph{Scientific Reports}, 3\penalty0 (1):\penalty0 2606, September
  2013.
\newblock ISSN 2045-2322.
\newblock \doi{10.1038/srep02606}.

\bibitem[van Haastert and De~Wit(1984)]{VanHaastert1984}
Peter J.~M. van Haastert and René J.~W. De~Wit.
\newblock Demonstration of receptor heterogeneity and affinity modulation by
  nonequilibrium binding experiments. {The} cell surface {cAMP} receptor of
  {Dictyostelium} discoideum.
\newblock \emph{Journal of Biological Chemistry}, 259\penalty0 (21):\penalty0
  13321--13328, 1984.

\bibitem[Zanotelli et~al.(2018)Zanotelli, Goldblatt, Miller, Bordeleau, Li,
  VanderBurgh, Lampi, King, and Reinhart-King]{Zanotelli2018}
Matthew~R. Zanotelli, Zachary~E. Goldblatt, Jordan~P. Miller, Francois
  Bordeleau, Jingjing Li, John~A. VanderBurgh, Mikayla~C. Lampi, Michael~R.
  King, and Cynthia~A. Reinhart-King.
\newblock Regulation of atp utilization during metastatic cell migration by
  collagen architecture.
\newblock \emph{Molecular Biology of the Cell}, 29\penalty0 (1):\penalty0 1--9,
  2018.
\newblock \doi{10.1091/mbc.E17-01-0041}.

\bibitem[Louis et~al.(1993)Louis, Saxe, and Kimmel]{Louis1993}
J.~M. Louis, C.~L. Saxe, and A.~R. Kimmel.
\newblock Two transmembrane signaling mechanisms control expression of the
  {cAMP} receptor gene {CAR1} during {Dictyostelium} development.
\newblock \emph{Proceedings of the National Academy of Sciences}, 90\penalty0
  (13):\penalty0 5969--5973, 1993.
\newblock \doi{10.1073/pnas.90.13.5969}.

\bibitem[Masaki et~al.(2013)Masaki, Fujimoto, Honda-Kitahara, Hada, and
  Sawai]{Masaki2013}
Noritaka Masaki, Koichi Fujimoto, Mai Honda-Kitahara, Emi Hada, and Satoshi
  Sawai.
\newblock Robustness of self-organizing chemoattractant field arising from
  precise pulse induction of its breakdown enzyme: a single-cell level analysis
  of {PDE} expression in {Dictyostelium}.
\newblock \emph{Biophysical Journal}, 104\penalty0 (5):\penalty0 1191--1202,
  2013.
\newblock ISSN 0006-3495.
\newblock \doi{10.1016/j.bpj.2013.01.023}.

\bibitem[Kirkegaard et~al.(2026)Kirkegaard, Alonso, and Endres]{Kirkegaard2026}
J.~B. Kirkegaard, A.~Alonso, and R.~G. Endres.
\newblock Limits of cell sensing in observer--absorber assemblies.
\newblock Manuscript submitted for publication, 2026.

\bibitem[Landau et~al.(1984)Landau, Lifshitz, and Pitaevskii]{Landau1984}
L.~D. Landau, E.~M. Lifshitz, and L.~P. Pitaevskii.
\newblock \emph{Electrodynamics of Continuous Media}, volume~8 of \emph{Course
  of Theoretical Physics}.
\newblock Pergamon Press, Oxford, 2nd edition, 1984.

\end{thebibliography}

\clearpage
\appendix
% appendix equations numbered A1, B2, ...
\renewcommand{\theequation}{\thesection\arabic{equation}}
\makeatletter\@addtoreset{equation}{section}\makeatother

\section*{Appendices}
These appendices give the derivations used in the main text.
The central object is the surface partial absorber, a cell of radius $a$ whose surface reactivity $\kappa$ imposes a Robin boundary condition on the exterior diffusion problem.
We derive the uptake rate, the monopole--dipole flux coefficients, the gradient-sensing uncertainty, and its counting form, in which the Damk\"ohler number is traded for the uptake fraction.
We then derive the entropy produced by transport, set out the units and parameter values used for the \textit{Dictyostelium} comparison, and obtain the uptake required to approach peak chemotactic alignment.
Two further sections ask how far the spherical geometry can be relaxed, and how this surface absorber relates to the volume partial absorber of the related observer--absorber study~\cite{Kirkegaard2026}.

Throughout, $D$ is the ligand diffusion constant, $c_0$ is the background concentration, $g$ is the magnitude of a shallow gradient directed, without loss of generality, along the $z$ axis, $T$ is the integration time, and entropy is measured in units of $k_B$.
The polar angle $\theta$ is measured from the gradient direction.

\section{Surface partial absorber and concentration counting}
\label{app:partial-absorber}

For a radially symmetric concentration field outside a spherical cell, the steady-state diffusion equation is
\begin{equation}
\frac{1}{r^2}\frac{d}{dr}\left(r^2\frac{dc}{dr}\right)=0,
\qquad r>a,
\end{equation}
with solution
\begin{equation}
c(r)=A+\frac{B}{r}.
\end{equation}
The far-field condition is $c(r\to\infty)=c_0$, so $A=c_0$.
The surface is partially absorbing, with Robin boundary condition
\begin{equation}
D\,\partial_r c\big|_{r=a}=\kappa c(a),
\label{eq:SI-robin-radial}
\end{equation}
where the inward flux density is written as $j=D\partial_r c|_{r=a}$.
Substitution of $c(r)=c_0+B/r$ into Eq.~\eqref{eq:SI-robin-radial} gives
\begin{equation}
D\left(-\frac{B}{a^2}\right)=\kappa\left(c_0+\frac{B}{a}\right),
\end{equation}
and hence
\begin{equation}
B=-\frac{\kappa a^2 c_0}{D+\kappa a}.
\end{equation}
The concentration profile is therefore
\begin{equation}
c(r)=c_0\left[1-\frac{\kappa a^2}{(D+\kappa a)r}\right].
\label{eq:SI-profile}
\end{equation}
The corresponding inward flux density is
\begin{equation}
j_0=D\partial_r c\big|_{r=a}=\frac{D\kappa c_0}{D+\kappa a}.
\label{eq:SI-j0-radial}
\end{equation}
Thus the total uptake rate is the Collins--Kimball result
\begin{equation}
J=4\pi a^2 j_0=4\pi D a c_0\frac{\kappa a}{D+\kappa a}.
\end{equation}
Introducing the surface Damk\"ohler number
\begin{equation}
\Da\equiv\frac{\kappa a}{D},
\end{equation}
the mean number of absorbed ligands in time $T$ is
\begin{equation}
\langle N\rangle=JT=4\pi D a c_0 T\frac{\Da}{1+\Da}.
\label{eq:SI-N}
\end{equation}
This interpolates between a perfect reflector, $\Da\to0$, and the perfect absorber, $\Da\to\infty$.
For concentration sensing based only on the total absorption count, Poisson statistics gives
\begin{equation}
\left(\frac{\delta c}{c_0}\right)^2
=\frac{1}{\langle N\rangle}
=\left(4\pi D a c_0T\frac{\Da}{1+\Da}\right)^{-1}.
\label{eq:SI-conc-error}
\end{equation}

\section{Monopole--dipole solution in a shallow gradient}
\label{app:monopole-dipole}

For a shallow far-field gradient along the $z$ axis,
\begin{equation}
c(r,\theta)\to c_0+gr\cos\theta
\qquad (r\to\infty),
\end{equation}
the axisymmetric solution of Laplace's equation can be expanded to first order in the gradient as
\begin{equation}
c(r,\theta)=c_0+\frac{B_0}{r}+\left(gr+\frac{B_1}{r^2}\right)\cos\theta.
\label{eq:SI-general-solution}
\end{equation}
The Robin boundary condition separates into monopole and dipole modes, a decomposition valid to first order in the shallow gradient.
The monopole mode is the radial problem of Sec.~\ref{app:partial-absorber}, returning $B_0=-\kappa a^2c_0/(D+\kappa a)$ and the flux density $j_0$ of Eq.~\eqref{eq:SI-j0-radial}.
For the dipole mode, Eq.~\eqref{eq:SI-general-solution} gives
\begin{equation}
D\left(g-\frac{2B_1}{a^3}\right)
=\kappa\left(ga+\frac{B_1}{a^2}\right),
\end{equation}
and hence
\begin{equation}
B_1=a^3g\frac{D-\kappa a}{2D+\kappa a}.
\end{equation}
The dipolar part of the inward surface flux is
\begin{equation}
j_1=D\left(g-\frac{2B_1}{a^3}\right)
=\frac{3D\kappa a g}{2D+\kappa a}.
\label{eq:SI-j1}
\end{equation}
Thus the angular flux profile is
\begin{equation}
j(\theta)=j_0+j_1\cos\theta,
\end{equation}
with
\begin{equation}
\frac{j_1}{j_0}
=\frac{3ag}{c_0}\frac{1+\Da}{2+\Da}.
\label{eq:SI-ratio}
\end{equation}
The bounded factor $(1+\Da)/(2+\Da)$ increases from $1/2$ to $1$ between the reflecting and absorbing limits.
The ratio alone is not the available information, however, because both $j_0$ and $j_1$ vanish as $\Da\to0$.

\section{Gradient estimator and uncertainty}
\label{app:gradient}

The cell does not observe the continuous flux profile $j(\theta)$, but a finite sequence of absorption events at angles $\theta_1,\dots,\theta_N$ drawn from $p(\theta)\propto j(\theta)\sin\theta$.
To first order in the gradient, the only statistic in that sample carrying directional information is the empirical mean $\overline{\cos\theta}$.
Its expectation is the flux-weighted angular moment
\begin{equation}
\langle\cos\theta\rangle_j
=\frac{\int j(\theta)\cos\theta\,dA}{\int j(\theta)\,dA},
\qquad
dA=2\pi a^2\sin\theta\,d\theta .
\end{equation}
Writing $u=\cos\theta$ turns both integrals into elementary ones over $u\in[-1,1]$, in each of which the term odd in $u$ drops out:
\begin{equation}
\int j\cos\theta\,dA=2\pi a^2\!\!\int_{-1}^{1}\!\!\left(j_0+j_1u\right)u\,du=\frac{4\pi a^2}{3}j_1 ,
\end{equation}
\begin{equation}
\int j\,dA=2\pi a^2\!\!\int_{-1}^{1}\!\!\left(j_0+j_1u\right)du=4\pi a^2j_0 .
\end{equation}
The monopole therefore sets the number of arrivals and the dipole their mean direction,
\begin{equation}
\langle\cos\theta\rangle_j=\frac{j_1}{3j_0},
\label{eq:SI-costheta}
\end{equation}
the factor of $3$ coming from $\int_{-1}^{1}u^2du=2/3$ against $\int_{-1}^{1}du=2$.
An unbiased gradient estimator can therefore be written as
\begin{equation}
\hat g=\frac{g}{\langle\cos\theta\rangle_j}\,\overline{\cos\theta}.
\end{equation}
In the shallow-gradient limit the angular distribution is nearly isotropic, so
\begin{equation}
\mathrm{Var}(\cos\theta)=\frac{1}{3}+O\!\left[(j_1/j_0)^2\right].
\end{equation}
The neglected terms are $3(ag/c_0)^2\left[(1+\Da)/(2+\Da)\right]^2$: at ${ag/c_0=0.1}$ they are $0.8\%$ for weak uptake and $3\%$ for a perfect absorber, and at ${ag/c_0=0.3}$, $7\%$ and $27\%$ respectively, so the expansion loses accuracy smoothly rather than at a threshold.
Across the six threshold conditions of Table~1 of~\citet{VanHaastert2007}, the fractional gradient across a cell of radius $a=5\;\upmu$m ranges from $0.7\%$ to $16\%$, so the shallow-gradient expansion holds over the entire data set, including at the steepest gradients.
With Poisson arrival statistics, the leading variance of the empirical angular mean is therefore
\begin{equation}
\mathrm{Var}(\overline{\cos\theta})\simeq\frac{1}{3\langle N\rangle}.
\end{equation}
Propagating this variance through the estimator gives
\begin{equation}
\left(\frac{\delta g}{g}\right)^2
=\frac{3}{\langle N\rangle}\left(\frac{j_0}{j_1}\right)^2.
\label{eq:SI-grad-rel}
\end{equation}
Substituting Eqs.~\eqref{eq:SI-N} and~\eqref{eq:SI-ratio} gives
\begin{equation}
\left(\frac{\delta g}{g}\right)^2
=\frac{c_0}{12\pi D a^3g^2T}
\frac{(2+\Da)^2}{\Da(1+\Da)}.
\label{eq:SI-grad-rel-Da}
\end{equation}
Equivalently, expressing the error in the natural gradient units $c_0/a$ removes the explicit shallow-gradient factor:
\begin{equation}
\left(\frac{\delta g}{c_0/a}\right)^2
=\frac{1}{12\pi D a c_0T}
\frac{(2+\Da)^2}{\Da(1+\Da)}.
\label{eq:SI-grad-natural-Da}
\end{equation}
In the perfect-absorber limit this reduces to
\begin{equation}
\left(\frac{\delta g}{c_0/a}\right)^2_{\Da\to\infty}
=\frac{1}{12\pi D a c_0T},
\end{equation}
which is the single-component gradient uncertainty.
For the three-dimensional gradient magnitude, the conventional total uncertainty is three times larger,
\begin{equation}
\left(\frac{\delta |\bm g|}{c_0/a}\right)^2_{\Da\to\infty}
=\frac{1}{4\pi D a c_0T},
\end{equation}
which is Eq.~(7) and the perfect-absorber column of Table~1 of Endres and Wingreen~\cite{Endres2008}.
Their Eq.~(6), $\langle(\delta c_z)^2\rangle=c_0/(12\pi Da^3T)$, is the same single-component statement.

Equation~\eqref{eq:SI-grad-rel} is what separates gradient sensing from concentration sensing, and its two factors are the two effects that limit the estimate.
The first, $1/\langle N\rangle$, is the usual counting noise, $\overline{\cos\theta}$ being a mean of $\langle N\rangle$ draws.
The second, $(j_0/j_1)^2$, is the geometric penalty for reading a small dipolar asymmetry off a dominant monopole flux, and it is large in relative units $\delta g/g$ when $ag/c_0\ll1$.
The perfect absorber improves both: it maximizes $\langle N\rangle$ at fixed $c_0$, and by Eq.~\eqref{eq:SI-ratio} it also maximizes the contrast $j_1/j_0$ through the bounded factor $(1+\Da)/(2+\Da)$.
The two enter Eq.~\eqref{eq:SI-counting} through the single combination $(2-\Phi)^2/\langle N\rangle$.

\section{Counting form and elimination of the Damk\"ohler number}
\label{app:counting}

It is convenient to trade the Damk\"ohler number for the uptake fraction
\begin{equation}
\Phi\equiv\frac{\langle N\rangle}{\langle N\rangle_{\max}}=\frac{\Da}{1+\Da},
\qquad
\langle N\rangle_{\max}=4\pi Dac_0T,
\label{eq:SI-x}
\end{equation}
which is the fraction of the diffusion-limited uptake the cell actually absorbs.
It is the natural variable here because it is a ratio of counts.
The Damk\"ohler number requires the surface reactivity $\kappa$, which is not observable; $\Phi$ is fixed by the uptake itself, and runs over the unit interval rather than the half-line.

Inverting gives $\Da=\Phi/(1-\Phi)$, so that
\begin{equation}
1+\Da=\frac{1}{1-\Phi},
\qquad
2+\Da=\frac{2-\Phi}{1-\Phi},
\qquad
\Da\left(1+\Da\right)=\frac{\Phi}{\left(1-\Phi\right)^2},
\end{equation}
and the Damk\"ohler-dependent factor of Eq.~\eqref{eq:SI-grad-natural-Da} becomes
\begin{equation}
\frac{(2+\Da)^2}{\Da(1+\Da)}
=\frac{(2-\Phi)^2}{(1-\Phi)^2}\cdot\frac{(1-\Phi)^2}{\Phi}
=\frac{(2-\Phi)^2}{\Phi}.
\end{equation}
The factors of $1-\Phi$ cancel, which is what removes the singular behaviour carried by $\Da\to\infty$.
Since $\langle N\rangle=\Phi\langle N\rangle_{\max}$, Eq.~\eqref{eq:SI-grad-natural-Da} collapses to
\begin{equation}
\left(\frac{\delta g}{c_0/a}\right)^2
=\frac{\left(2-\Phi\right)^2}{3\,\langle N\rangle}.
\label{eq:SI-counting}
\end{equation}
Two features of Eq.~\eqref{eq:SI-counting} are worth separating.
The entire dependence on the boundary condition has collapsed into $(2-\Phi)^2$, a factor bounded between $4$ and $1$; everything else is the counting noise $1/\langle N\rangle$ familiar from concentration sensing.
And nothing thermodynamic has entered, so the relation ties an error to a count and to nothing else.

\section{Entropy production}
\label{app:entropy-derivation}

Each absorbed ligand is internalized or degraded through a process producing entropy $\Delta s$.
In steady state these events are independent and identical, so the entropy produced over the measurement is the per-event cost times the number of events,
\begin{equation}
\Sigma=\langle N\rangle\Delta s.
\label{eq:SI-Sigma}
\end{equation}
The two contributions to $\Delta s$, the entropy $\Delta s_{\mathrm{tr}}$ produced in delivering the ligand to the surface and the biochemical cost $\Delta s_0$ of registering and disposing of it, are established in turn below and then combined with Eq.~\eqref{eq:SI-counting}.

\subsection{Entropy produced by transport}
\label{app:transport}

The entropy produced in delivering a ligand to the surface is fixed by the concentration profile.
The local entropy production rate of a diffusive flux is that flux worked against the gradient of its chemical potential,
\begin{equation}
\sigma=-\frac{\bm{J}_c\cdot\nabla\mu}{k_BT},
\qquad
\bm{J}_c=-D\nabla c .
\end{equation}
For an ideal solution $\mu=\mu_0+k_BT\ln c$, so that $\nabla\mu=k_BT\nabla c/c$ and
\begin{equation}
\sigma=\frac{D\abs{\nabla c}^2}{c} .
\label{eq:SI-sigma}
\end{equation}
Take the radial profile of Eq.~\eqref{eq:SI-profile}, written as $c(r)=c_0\left(1-\Phi a/r\right)$ with $\Phi=\Da/(1+\Da)$ the uptake fraction, for which the steady-state current is $J=4\pi Dac_0\Phi$.
Differentiating and substituting into Eq.~\eqref{eq:SI-sigma},
\begin{equation}
\nabla c=\frac{c_0\Phi a}{r^2}\,\hat{\bm{r}},
\qquad
\sigma=\frac{Dc_0\Phi^2a^2}{r^3\left(r-\Phi a\right)} ,
\end{equation}
and integrating over the exterior with $u=r/a$,
\begin{equation}
\dot\Sigma_{\mathrm{tr}}
=\int_a^\infty \sigma\,4\pi r^2dr
=4\pi Dac_0\Phi^2\int_1^\infty\frac{du}{u\left(u-\Phi\right)} .
\end{equation}
Partial fractions give ${\left[u(u-\Phi)\right]^{-1}=\Phi^{-1}\left[(u-\Phi)^{-1}-u^{-1}\right]}$, so the integral is
\begin{equation}
\int_1^\infty\frac{du}{u\left(u-\Phi\right)}
=\frac{1}{\Phi}\left[\ln\frac{u-\Phi}{u}\right]_1^\infty
=-\frac{\ln\left(1-\Phi\right)}{\Phi} ,
\end{equation}
the upper limit vanishing because $\ln[(u-\Phi)/u]\to0$ as $u\to\infty$.
Hence
\[
\dot\Sigma_{\mathrm{tr}}=4\pi Dac_0\,\Phi\left[-\ln(1-\Phi)\right],
\]
and dividing by the current $J$ gives the entropy produced per internalized molecule,
\begin{equation}
\Delta s_{\mathrm{tr}}=\frac{\dot\Sigma_{\mathrm{tr}}}{J}=-\ln\left(1-\Phi\right)=\ln\left(1+\Da\right)=\ln\frac{c_0}{c(a)},
\label{eq:SI-ds-transport}
\end{equation}
which is exactly the chemical-potential drop $\Delta\mu/k_BT$ between the bulk and the surface.
It vanishes at the reflector and diverges logarithmically at the perfect absorber, echoing the unbounded entropy production that ideal maximum-likelihood estimation demands of receptor kinetics~\cite{Lang2014}, here from the boundary condition instead.

The result is shape-independent.
For an arbitrary absorbing body let $\phi$ solve Laplace's equation with $\phi=1$ on the surface and $\phi\to0$ at infinity, so that $c=c_0\left(1-\Phi\phi\right)$.
This profile holds the surface at the uniform concentration $c_0(1-\Phi)$, which on a non-spherical body requires the reactivity to follow the local perfect-absorber flux; on the sphere it is automatic.
Using the co-area formula and the fact that $\int\abs{\nabla\phi}\,dS=4\pi C$ is the same on every level set, with $C$ the capacitance, the radial integral above is replaced by an identical one-dimensional integral in the level-set variable.
The capacitance cancels between $\dot\Sigma_{\mathrm{tr}}$ and $J$, leaving Eq.~\eqref{eq:SI-ds-transport} unchanged for any such absorber.

The calculation retains the monopole only, the dipole correction being of order ${(ag/c_0)^2}$, and assumes an ideal solution, ${\mu=\mu_0+k_BT\ln c}$.

\subsection{Interpretation of \texorpdfstring{$\Delta s_0$}{Delta s0}}
\label{app:microscopic}

The Robin boundary condition fixes the coarse-grained kinetic accessibility of the surface through $\kappa$, but it does not by itself fix the thermodynamic affinity of the microscopic internalization process.
For example, a minimal reversible receptor cycle may be written schematically as
\begin{equation}
R_u+L
\xrightleftharpoons[k_{\mathrm{off}}]{k_{\mathrm{on}}c(a)}
R_b
\xrightleftharpoons[k_{\mathrm{rec}}]{k_{\mathrm{int}}}
R_i,
\label{eq:SI-micro-cycle}
\end{equation}
where $R_u$, $R_b$, and $R_i$ denote unbound, ligand-bound, and internalized receptor states.
The entropy produced per completed uptake cycle is the cycle affinity
\begin{equation}
\Delta s=
\ln\left(\frac{k_{\mathrm{on}}c(a)\,k_{\mathrm{int}}}{k_{\mathrm{off}}\,k_{\mathrm{rec}}}\right),
\label{eq:SI-affinity}
\end{equation}
provided the underlying rates are resolved.
In a dilute unsaturated limit, the same microscopic rates generate an effective surface reactivity of the form
\begin{equation}
\kappa_{\mathrm{eff}}
\simeq
\frac{R_{\mathrm{tot}}}{4\pi a^2}
\frac{k_{\mathrm{on}}k_{\mathrm{int}}}{k_{\mathrm{off}}+k_{\mathrm{int}}},
\label{eq:SI-kappa-eff}
\end{equation}
up to model-dependent corrections from receptor recycling and saturation.
Equations~\eqref{eq:SI-affinity} and~\eqref{eq:SI-kappa-eff} show why $\kappa$ and $\Delta s$ need not be in one-to-one correspondence after coarse graining, since different microscopic cycles can have the same effective absorption strength but different thermodynamic driving.
The affinity in Eq.~\eqref{eq:SI-affinity} diverges as $k_{\mathrm{rec}}\to0$.
This limit is an idealization rather than a physical cost: internalization is a multistep process with a finite reverse rate, which bounds $\Delta s_0$.

The magnitude of $\Delta s_0$ is architecture-dependent and spans three orders of magnitude across known systems.
Optimal allocation of sensing resources gives $\sim\!1\,k_B$ per independent sample~\cite{Govern2014a, Govern2014b, TenWolde2016}.
That accounting is per receptor-occupancy sample rather than per absorbed molecule, so it is not directly the $\Delta s$ of Eq.~\eqref{eq:SI-Sigma}.
At the opposite extreme, a ligand-bound G-protein-coupled receptor catalyses nucleotide exchange on many G proteins~\cite{Janetopoulos2001}, each hydrolysis costing $\sim\!20\,k_B$~\cite{Milo2015}, which places $\Delta s_0$ at $10^2$--$10^3\,k_B$ per detection event.
Nor need the chemistry be endergonic. In \textit{Dictyostelium} the cAMP receptor cAR1 is not internalized, and external cAMP is instead hydrolysed by membrane phosphodiesterase~\cite{Endres2008}, which releases free energy rather than consuming it.
The steps that register and dispose of a captured ligand nevertheless run in series and carry the same flux in steady state, so the entropy produced per completed cycle is the cycle affinity, which carries the same sign as that flux.
Sustaining a net uptake current therefore requires ${\Delta s_0\geq0}$ whatever the sign of the free-energy change of any individual step, and an exergonic step contributes more to the affinity rather than less.

The cycle of Eq.~\eqref{eq:SI-micro-cycle} is a representative model rather than a realistic one.
In a chemotactic eukaryote a captured ligand is registered by a receptor cascade of many coupled cycles, and $\Delta s_0$ lumps all of them into a single per-event cost.
It should therefore be read as an order of magnitude set by the architecture, not as a quantity the minimal cycle predicts, which is why the main text leaves it free and states its results as counts wherever they can be.

\subsection{The energy--accuracy relation}

Eliminating $\langle N\rangle$ between Eqs.~\eqref{eq:SI-counting} and~\eqref{eq:SI-Sigma}, with $\Delta s=\Delta s_0-\ln(1-\Phi)$ the entropy per absorbed molecule, gives the energy--accuracy relation of the main text as an identity,
\begin{equation}
\left(\frac{\delta g}{c_0/a}\right)^2
=\frac{(2-\Phi)^2\left[\Delta s_0-\ln(1-\Phi)\right]}{3\Sigma}.
\label{eq:SI-tradeoff}
\end{equation}
Equation~\eqref{eq:SI-tradeoff} separates the scalar counting contribution, proportional to $1/\Sigma$, from the dipole-contrast factor $(2-\Phi)^2$.
It is a change of variable rather than an independent statement, since $\Sigma$ and $\langle N\rangle$ are proportional and Eq.~\eqref{eq:SI-counting} already contains the physics.
What the change of variable buys is the constant of proportionality: $\Delta s$ is architecture-dependent and known only to within orders of magnitude, so a relation written in $\Sigma$ inherits that spread, whereas one written in $\langle N\rangle$ does not.
This is why the results of the main text are quoted as counts, and why the bound below is the useful form of Eq.~\eqref{eq:SI-tradeoff}: an inequality survives an uncertain prefactor where an equality does not.

Discarding the transport term, $-\ln(1-\Phi)\geq0$, gives
\begin{equation}
\left(\frac{\delta g}{c_0/a}\right)^2
\geq\frac{(2-\Phi)^2\Delta s_0}{3\Sigma}
\label{eq:SI-bound}
\end{equation}
at every uptake fraction.
What is discarded is the transport entropy, set by the cell's uptake but dissipated in the solution outside it, and negligible away from the diffusion limit; the surviving $\Delta s_0$ is the biochemical share.
Since $\Delta s_0$ omits receptor synthesis and the standing cost of the signaling network, it understates what the cell spends, and the bound is a conservative statement about metabolic cost.

\subsection{Relation to the thermodynamic uncertainty relation}
\label{app:tur}

The scalar thermodynamic uncertainty relation (TUR) bounds the precision of a time-integrated current by the entropy produced by the same microscopic process~\cite{Barato2015, Horowitz2020}.
For the total absorption count $N$, the scalar-current form constrains the counting part of the sensing problem.
For concentration sensing, the estimator is essentially the total count, so the uncertainty follows $1/\langle N\rangle$.
For gradient sensing, the estimator is not just the total count, but an angular moment of the absorbed flux.
Thus the scalar-current TUR does not by itself fix the gradient uncertainty; the extra factor in Eq.~\eqref{eq:SI-tradeoff} records the angular information needed to infer the dipole.

Equation~\eqref{eq:SI-tradeoff} also cannot be inverted into a bound on $\Delta s$, because a gradient estimate has no intrinsic relative variance.
The normalization of $\delta g$ is a choice: measuring it against $c_0/a$, as in Eq.~\eqref{eq:SI-counting}, and against $g$ itself differ by the factor $(ag/c_0)^{-2}$, so the same absorption events would return different apparent floors.
The constraint on $\Delta s$ used in the main text comes instead from transport, Eq.~\eqref{eq:SI-ds-transport}, where the entropy produced in the solution is fixed by the concentration profile rather than inferred from an inequality.

\section{Units and parameter values}
\label{app:units}

Throughout, $c_0$ is a number density, which is what makes the group $A\equiv3\pi Da^3T\,g^2/c_0$ of Eq.~\eqref{eq:SI-z-x} dimensionless.
With $[D]=\upmu$m$^2$/s, $[a]=\upmu$m and $[T]=$~s, $Da^3T$ has units of $\upmu$m$^5$ and $g^2/c_0$ is in $\upmu$m$^{-5}$.
Experimental gradients are quoted in molar units, converted with $1$~L~$=10^{15}\,\upmu$m$^3$ as
\begin{equation}
1\ \mathrm{nM}=\frac{6.022\times10^{23}\ \mathrm{L}^{-1}\times10^{-9}}{10^{15}\ \upmu\mathrm{m}^3\,\mathrm{L}^{-1}}=0.6022\ \upmu\mathrm{m}^{-3},
\label{eq:SI-molar}
\end{equation}
so a value of $g^2/c_0$ in nM\,$\upmu$m$^{-2}$ carries a factor $0.6022$ into Eq.~\eqref{eq:SI-z-x}.

For the \textit{Dictyostelium} comparison of Fig.~\ref{fig:chemo} we take $D=300\;\upmu$m$^2$/s for cAMP, $a=5\;\upmu$m, and $T=4$~s, so
\begin{equation}
Da^3T=300\times5^3\times4=1.5\times10^{5}\ \upmu\mathrm{m}^5 .
\end{equation}
The abscissa of Fig.~\ref{fig:chemo} is $g^2/c_0$ in nM\,$\upmu$m$^{-2}$, obtained from Table~1 of~\citet{VanHaastert2007}, which lists the background concentration $C$ and spatial gradient $\nabla C$ at the threshold of chemotaxis for each assay; for example, at $700\;\upmu$m from a $100\;\upmu$M pipette, $C=7.14$~nM and $\nabla C=10.20$~pM$/\upmu$m, giving $g^2/c_0=1.46\times10^{-5}$~nM\,$\upmu$m$^{-2}$.
The value of $T$ is not a free parameter of the transport problem but a property of the signaling network: \citet{VanHaastert2007} infer a second-messenger lifetime and time-averaging window of $2$--$8$~s from their receptor-noise analysis, and we take $T=4$~s from within it.
Varying $T$ across that window changes the fitted uptake fraction $\Phi$ by a factor of four but leaves the absolute uptake $\langle N\rangle=\Phi\,4\pi Dac_0T$ at the threshold condition constant to within $1\%$.

\section{Chemotactic index and cost-optimal uptake}
\label{app:ci}

The main text propagates the sensing uncertainty to the chemotactic index of~\citet{Endres2008}, used in the same form in~\cite{Alonso2025a},
\begin{equation}
\mathrm{CI}(z)=\sqrt{\frac{\pi z}{2}}e^{-z}\left[I_0(z)+I_1(z)\right],
\label{eq:SI-CI}
\end{equation}
where $I_0$ and $I_1$ are modified Bessel functions.
Using Eq.~\eqref{eq:SI-grad-natural-Da}, and then eliminating $\Da$ in favor of the uptake fraction $\Phi$, the argument can be written as
\begin{equation}
z=A\frac{\Phi}{(2-\Phi)^2},
\qquad
A\equiv 3\pi D a^3T\frac{g^2}{c_0}.
\label{eq:SI-z-x}
\end{equation}
Thus the behavioral response depends on the environment and integration time only through the dimensionless group $A$, and on surface absorption only through $\Phi$.
At $\Phi\to1$ the argument reduces to $z\to A$, which is the variable $y=(c_z^0)^2/(4\sigma_z^2)$ of their Eq.~(23), so Eq.~\eqref{eq:SI-CI} is the perfect-absorber index and $A$ is their $3\pi Da^3Tg^2/c_0$.
See Sec.~\ref{app:units} for the units of $A$ and the parameter values used.

We define $\Phi^*$ as the smallest uptake fraction that reaches a fraction $\alpha$ of the perfect-absorber chemotactic index,
\begin{equation}
\mathrm{CI}\!\left(A\frac{\Phi^*}{(2-\Phi^*)^2}\right)=\alpha\,\mathrm{CI}(A).
\end{equation}
Let
\begin{equation}
z_\alpha=\mathrm{CI}^{-1}\!\left[\alpha\,\mathrm{CI}(A)\right],
\qquad
\zeta=\frac{z_\alpha}{A}.
\end{equation}
Then $\Phi^*$ solves $\Phi^*/(2-\Phi^*)^2=\zeta$, a quadratic whose physical root is
\begin{equation}
\Phi^*
=\frac{1+4\zeta-\sqrt{1+8\zeta}}{2\zeta},
\label{eq:SI-xstar}
\end{equation}
the minus branch being the one lying in $0<\Phi^*<1$ for small $\zeta$.
In the strong-signal regime, $A\gg1$, $\mathrm{CI}(A)\to1$ and
\begin{equation}
\Phi^*
\simeq \frac{4\mathrm{CI}^{-1}(\alpha)}{A}.
\label{eq:SI-xstar-asymp}
\end{equation}
For $\alpha=0.9$, $\mathrm{CI}^{-1}(0.9)=1.497$ and $4\mathrm{CI}^{-1}(0.9)=5.99\simeq6$, giving the asymptote used in the main text.
Multiplying Eq.~\eqref{eq:SI-xstar-asymp} by $\langle N\rangle_{\max}=4\pi Dac_0T$ gives the cost-optimal uptake in absorbed-ligand units,
\begin{equation}
N^*=\Phi^*\langle N\rangle_{\max}
\simeq
\frac{16}{3}\mathrm{CI}^{-1}(\alpha)
\left(\frac{c_0}{ag}\right)^2 ,
\label{eq:SI-absolute-cost}
\end{equation}
with prefactor $(16/3)\,\mathrm{CI}^{-1}(0.9)=7.98$.
The fraction $\Phi^*$ depends on $A$, and hence on $D$ and $T$.
The count in Eq.~\eqref{eq:SI-absolute-cost} does not, since $D$ and $T$ cancel between $A$ and $\langle N\rangle_{\max}$, leaving $N^*$ a function of $ag/c_0$ alone.
The comparison with experiment in Sec.~\ref{app:units} is therefore insensitive to the choice of integration time.
The asymptotic count of Eq.~\eqref{eq:SI-absolute-cost} is within $7\%$ of the exact one once $A\gtrsim10^2$, but overstates it by $64\%$ at $A=10$, where $\mathrm{CI}(A)$ has not yet saturated.

The time $T$ is the sensory integration time, distinct from the motility or decision-cycle time $T_d$ over which the cell reorients.
The chemotactic index of Eq.~\eqref{eq:SI-CI} is a purely angular alignment measure: it does not depend on speed, drift velocity, or step length, and hence not on $T_d$.
For a freely swimming bacterium $T$ is bounded above by rotational diffusion and is shorter than for a crawling eukaryote, while $A\propto Da^3T$ falls faster still with the smaller cell radius; by Eq.~\eqref{eq:SI-xstar-asymp} the required $\Phi^*$ is correspondingly larger, so bacteria must work nearer the diffusion limit to reach the same alignment.

\section{Cell shape and the substrate}
\label{app:geometry}

The main text models the cell as a free sphere, the geometry of~\citet{Berg1977}, \citet{Endres2008} and~\citet{Kaizu2014a}.
Real chemotactic cells are neither free nor spherical.
\textit{Dictyostelium} crawls on a surface and is somewhat flattened.
Neither departure touches the structure of Eqs.~\eqref{eq:SI-counting}--\eqref{eq:SI-bound}.
The subsections below treat the substrate, the total uptake, and the directional response in turn, comparing bodies of equal cell volume throughout.

\subsection{An adhered cell is exactly half of its reflected body}

An impermeable substrate imposes no ligand flux through the plane, ${\partial_zc=0}$.
For a gradient directed parallel to the plane the far field has no $z$ dependence at all, so reflecting the exterior field through the plane produces a solution that is even in $z$ and satisfies the no-flux condition automatically.
The image construction is therefore exact, and the adhered cell sees precisely the field of its free, doubled body, restricted to the upper half space.
By that symmetry the cell collects exactly half of the doubled body's uptake and half of its dipole moment.
The estimator reads direction from the angular distribution of arrivals through the moment ${\langle\cos\theta\rangle_j=j_1/(3j_0)}$ of Eq.~\eqref{eq:SI-costheta}, and halving both leaves that ratio alone, so every counting relation carries over.
Only the ligand budget changes.
A hemispherical cell absorbs
\begin{equation}
\frac{\langle N\rangle_{\mathrm{adhered}}}{\langle N\rangle_{\mathrm{free}}}
=\tfrac12\,2^{1/3}=2^{-2/3}=0.630
\label{eq:SI-adhered}
\end{equation}
of the equal-volume free sphere, the $2^{1/3}$ being the doubled body's radius.

\subsection{Deformation changes total uptake weakly}

Steady diffusion to a perfect absorber obeys Laplace's equation with $c=0$ on the surface, which is the electrostatics of a grounded conductor with the concentration in place of the potential.
The total diffusive current then plays the role of the conductor's charge, so the uptake rate is set by the electrostatic capacitance of the body, ${\langle N\rangle_{\max}=4\pi Dc_0TC}$, with $C=a$ for a sphere.
Capacitance grows with linear extent rather than with volume, and at fixed volume the sphere, being the most compact shape, \emph{minimizes} it, so any deformation raises the uptake.
The effect is weak.
The standard closed form for a prolate spheroid~\cite[\S4]{Landau1984} gives $C/R=1.04$, $1.11$ and $1.18$ at aspect ratios $2$, $3$ and $4$, with $R$ the equal-volume sphere radius.
Taken with the adhesion factor of Eq.~\eqref{eq:SI-adhered}, which acts the other way and is the larger of the two, a $3\!:\!1$ spread adhered cell absorbs ${0.630\times1.11=0.70}$ of what a free sphere of equal cell volume would.
Deformation is thus an under-$20\%$ rescaling of $\langle N\rangle_{\max}$, absorbed into the fitted $Da^3T$, with no effect on the uptake fraction $\Phi$ or on any counting relation.

\subsection{Elongation makes sensing anisotropic}

The capacitance fixes only the total uptake.
The direction information sits in the dipolar part of the flux, and there elongation acts differently along and across the long axis.

Two quantities carry the shape dependence of the counting relation.
The first is the dipole contrast.
The $2$ of Eq.~\eqref{eq:SI-counting} is, for the sphere, the ratio of the surface dipole of a perfect absorber to that of a reflector of the same body, and on a spheroid it splits into a pair of indices, $\nu_\parallel$ and $\nu_\perp$, for gradients along and across the symmetry axis.
A body drawn out along the gradient hardly perturbs the iso-concentration surfaces, so absorption barely sharpens the front--back contrast and $\nu_\parallel$ falls below $2$, whereas across the short axis the screening is stronger and $\nu_\perp$ rises above $2$.
The second is the lever arm.
The $3a^2$ of the spherical relation is the flux-weighted second moment $\mathsf{M}$ of the arrival positions, and arrivals spread over a longer stretch of surface carry more directional information per molecule, so $\mathsf{M}$ grows along the long axis.
In the slender limit $\nu_\parallel\to1$, $\nu_\perp$ grows only logarithmically, and $\mathsf{M}_\parallel$ approaches $L^2/12$, the position variance on a rod of length $L$, so a slender cell measures the gradient against its own length.

Solving the exterior problem in prolate spheroidal coordinates, the route by which \citet[\S4]{Landau1984} obtain the ellipsoidal depolarization coefficients, gives ${\nu_\parallel=1.81}$ and ${\nu_\perp=2.16}$ at aspect ratio $2$, about that of \textit{E.\ coli} ($1.75$ and $2.24$ at $2.5$), straddling the spherical $\nu=2$.
The required count scales as $\nu^2/\mathsf{M}$, and at fixed cell volume it is $0.61$ of the equal-volume sphere's along the long axis and $1.23$ across it ($0.50$ and $1.33$ at $2.5$).
The requirement therefore stays within about a factor of two of the sphere up to $2.5\!:\!1$.
Since the uptake itself is orientation-independent, the alignment achieved at a given uptake depends on the angle between the gradient and the long axis, a testable prediction we do not pursue here.
Carrying $\nu$ and $\mathsf{M}$ through the paper would replace two numbers by functions of the aspect ratio without changing what any relation says, so the main text keeps the sphere.

\section{Relation to the volume partial absorber}
\label{app:relation-volume}

The surface partial absorber used here is not the same microscopic model as the volume partial absorber considered in the related observer--absorber study~\cite{Kirkegaard2026}.
In the present work, imperfect uptake is represented by a partially reactive surface, so the natural control parameter is
\begin{equation}
\Da=\frac{\kappa a}{D},
\end{equation}
and the uptake fraction relative to a perfect absorber is
\begin{equation}
\Phi_{\mathrm{surf}}(\Da)=\frac{\langle N\rangle}{4\pi Dac_0T}
=\frac{\Da}{1+\Da},
\label{eq:SI-Phi-surf}
\end{equation}
which is the $\Phi$ of Eq.~\eqref{eq:SI-x}.
The subscript distinguishes it from the volume absorber below, whose uptake fraction depends differently on its own control parameter.
In a volume partial absorber, ligand is removed throughout the cell interior rather than only at its boundary.
For a uniform first-order volumetric removal rate, the natural control parameter is
\begin{equation}
\Davi=\frac{\kappa_v a^2}{D},
\end{equation}
and the corresponding uptake fraction can be written as
\begin{equation}
\Phi_{\mathrm{vol}}(\Davi)
=1-\frac{\tanh\sqrt{\Davi}}{\sqrt{\Davi}}.
\label{eq:SI-Phi-vol}
\end{equation}
Both models have the same perfect-reflector and perfect-absorber endpoints,
\begin{equation}
\Phi\to0 \quad\text{as uptake vanishes},
\qquad
\Phi\to1 \quad\text{in the perfect-absorber limit}.
\end{equation}
For concentration sensing, both can therefore be written in the common counting-noise form
\begin{equation}
\left(\frac{\delta c}{c_0}\right)^2
=\frac{1}{4\pi Dac_0T}\frac{1}{\Phi}.
\label{eq:SI-unified-Phi}
\end{equation}
The difference is the map from the microscopic uptake strength to the uptake fraction $\Phi$.
For gradient sensing, this distinction matters more because the monopole and dipole modes need not be modified by uptake in exactly the same way.
The present paper focuses on the surface Robin problem, where that angular dependence produces the factor $(1+\Da)/(2+\Da)$ in Eq.~\eqref{eq:SI-ratio}.
The observer--absorber work instead studies volume absorption and collective depletion effects, where absorption can both suppress rebinding and reduce the ligand field available to neighboring cells.
Thus the two partial-absorber descriptions are complementary: they share the same perfect-absorber endpoint and the same concentration-counting structure when expressed through $\Phi$, but they describe different physical mechanisms away from that limit.

\end{document}